\documentclass{ws-ijmpa}

\usepackage{times}
\usepackage{psfig}
\usepackage{epsfig}
\usepackage{graphicx}

\def\be{\begin{equation}}

\def\ee{\end{equation}}

\begin{document}

\markboth{Mosquera Cuesta }{ Einstein's gravitational lensing and nonlinear electrodynamics  }

\catchline{}{}{}{}{}

\title{Einstein's gravitational lensing and nonlinear electrodynamics }

\author{Herman J. Mosquera Cuesta\footnote{ Abdus Salam International 
Centre for Theoretical Physics, Strada Costiera 11,  Miramare 34014, 
Trieste, Italy }}

\address{Instituto de Cosmologia, Relatividade e Astrof\'{\i}sica 
(ICRA-BR), Centro Brasileiro de Pesquisas F\'\i sicas \\ Rua Dr. Xavier 
Sigaud 150, CEP 22290-180, Rio de Janeiro, RJ, Brazil; hermanjc@cbpf.br}

\author{Jos\'e A. de Freitas Pacheco}

\address{Observatoire de la C\^ote d'Azur, BP 4229, F-06304, Nice Cedex 4, 
France}

\author{Jos\'e M. Salim}

\address{Instituto de Cosmologia, Relatividade e Astrof\'{\i}sica 
(ICRA-BR), Centro Brasileiro de Pesquisas F\'\i sicas \\ Rua Dr. Xavier 
Sigaud 150, CEP 22290-180, Rio de Janeiro, RJ, Brazil}


\maketitle



\begin{abstract}
Einstein (1936) predicted the phenomenon presently known as
gravitational lensing (GL).  A prime feature of GL is the
magnification, because of the gravitational field, of the star visible
surface as seen from a distant observer.  We show here that nonlinear
electrodynamics (NLED) modifies in a fundamental basis Einstein's
general relativistic (GR) original derivation. The effect becomes
apparent by studying the light propagation from a strongly magnetic
($B$) pulsar (SMP). Unlike its GR counterpart, the photon dynamics in
NLED leads to a new effective GL, which depends also on the $B$-field
permeating the pulsar. The apparent radius of a SMP appears then  
unexpectedly diminished, by a large factor, as compared to the classical 
Einstein's prediction. This may prove very crucial in determining physical
properties of high $B$-field stars from their X-ray emission.
\vskip 0.5 truecm 
\centerline{Received{(07/04/2004)}   --   Revised{(16/05/2005)}}
\end{abstract}

\keywords{Gravitational lensing --- Relativity --- stars: neutron --- stars: 
magnetic fields --- nonlinear electrodynamics}



\section{INTRODUCTION}

Einstein's general theory of relativity (GTR) has proved to be one of
the most successful physical theories ever formulated. As a theory of
the gravitational interaction, it has a number of predictions
definitely corroborated by experiments or astronomical observations,
which makes it the correct gravity theory here-to-fore. After
readdressing an earlier study, Einstein (1936) \cite{Einstein1936}
came up with
the prediction of the gravitational lensing effect. The idea is that the
gravitational field produced by a massive astrophysical object, the Sun
for instance, can act as a convergent lens able to deviate light
rays flying-by. Observations of far away quasars and galaxies and more
familiar total solar eclipses confirm the reality of this phenomenon. More
than half of the star's surface may be seen by a distant observer. This
means that a photon emitted at a given colatitude on the star's surface,
reaching the observer at infinity, must be emitted at a smaller angle
with respect to the normal at that point. Because of this effect,
a distant compact star must then appear to any observer larger then it
actually is. The relation between the apparent radius $R_{\infty}$ of a
spherical (non-rotating) star, as seen by a distant observer, and its
physical radius $R$ is

\begin{equation}
R_{\infty} = \frac{R} {\left(1 - \frac{R_{S}}{R}\right)^{1/2} } \; . 
\label{R-MAX}
\end{equation}

This relation is obtained from the photon trajectory in polar coordinates
($r, \theta$) by using a Schwarzschild metric

\begin{equation} 
ds^2 =  \left(1 - \frac{ R_{S} }{r}\right) dt^2 - \frac{dr^2}{ \left(1 - 
\frac{ R_{S} }{r}\right) } - r^2(d\theta^2 + \sin^2 \theta d\phi^2) \; . 
\label{SCHW} 
\end{equation}

Here $R_{S} = 2 G M/c^2$ is the Schwarzschild radius of the star of
mass $M$. As is clear from this line-element no effects from physical
fields other than the gravitational have been taken into account in
prescribing the star apparent size. This {\it paper} shows that
the phenomenon changes in a fundamental fashion if NLED is called into
play. The analysis below, based on well-known nonlinear Lagrangeans, as 
the Heisenberg-Euler (H-E, \cite{H-E}) and the exact Born \& Infeld (1934) 
(B-I \cite{B-I}), proves that when NLED is included to describe the photon 
dynamics, the trajectory depends on the background $B$-field pervading 
the star. We stress from the very beginning that the results below are 
obtained upon an idealization of the $B$-field structure. In a real star, 
the $B$-field is neither purely radial nor constant, so that the apparent 
radius hardly goes to zero at high $B$.

We stress in passing that this theory {\it is not a new theory of
gravitation}.  In our formalism Einstein's general relativity remains
definitively unaltered.  Only the photons propagating out of the
ultramagnetized neutron stars ``see" the effective metric that appears
once one incorporates NLED into a self-consistent physical description
of light propagation from those stars. That NLED is a
theory already tested in a number of experiments is demonstrated by the
laboratory experiment described by  Burke,
Field, Horton-Smith, et al. (1997)  \cite{BFH1997}, as quoted below.
Our theory attempts to call to the attention of the astrophysics
community the fundamental changes that should appear in determining
physical properties as radius, mass, and more crucial: the equation of
state, of neutron star pulsars endowed with extremely strong magnetic
fields, since a property as the star radius (which can be estimated
from the x-ray variability of very short period of the star emission
(light curve), or by measuring the gravitational redshift of absorption
or emission lines received from the star surface, or etc.) can say too 
much about the stellar structure and its equation of state. A fundamental
property that encodes basic information of the physical constituents of 
the neutron star matter.


\section{ Apparent radius in NLED }

According to quantum electrodynamics 
(QED: see Delphenich 2003 \cite{Delphenich2003} for a complete review on 
NLED and QED) a vacuum has nonlinear properties \cite{H-E,Schwinger51},
which affect the photon propagation. A noticeable advance in the
realization of this theoretical prediction has been provided by  Burke,
Field, Horton-Smith, et al. (1997) \cite{BFH1997}, who  demonstrated
experimentally that the inelastic scattering of laser photons by
gamma-rays is definitely a nonlinear phenomenon. The propagation of
photons in NLED has been examined by several authors (Bialynicka-Birula 
\& Bialynicka-Birula 1970; Salazar et al. 1989; Dietrich \& Gies 1998; 
De Lorenci 2000; Denisov \& Svertilov 2003; Alarcon 1981; Garcia \& 
Plebanski 1989; Boillat 1970; Vazquez et al. 1969; Torrence 1984; Garcia 1984)
\cite{Bialynicka-Birula1970,garcia89,dietrich98,vittorio,Denisov,plebanski81,plebanski-garcia89,boillat70,vazquez69,torrence84,garcia84}.  In
the geometric optics approximation, it was shown by Novello et {al.} (2000)
\cite{Novelloal2000}; and Novello \& Salim (2001) \cite{NS2001}, that when 
the photon propagation is identified with the propagation of
discontinuities of the EM field in a nonlinear regime, a remarkable
feature affloats:  the discontinuities propagate along null geodesics
of an {\it effective} geometry which depends on the EM field on the
background.  This means that the NLED interaction can be geometrized,
in analogy to gravity in GTR. A key outcome of this formalism is 
introduced in this {\it paper}.


\subsection{Euler-Heisenberg approach}

The H-E \cite{H-E} Lagrangean

\begin{equation}
L(F,G) = - \frac{1}{4}F + \frac{\mu}{4}\left(F^2 +
\frac{7}{4}G^2\right)\; , \label{H-E}
\end{equation}

where $\mu = \frac{2\alpha^2}{45m_e^4}$ is a quantum parameter, is a gauge
invariance description of the photon propagation in NLED that uses two
invariants

\begin{equation}
F = F_{\mu\nu}F^{\mu\nu}, \hskip 0.5 truecm G = F^{*}_{\mu\nu}F^{\mu\nu} 
\equiv \frac{1}{2}\eta_{\mu\nu}\mbox{}^{\alpha\beta}F_{\alpha\beta}F^{\mu\nu},
\label{eq3}
\end{equation}

constructed upon the Maxwell EM tensor and its dual \footnote{The
attentive reader must notice that this first order approximation is
valid only for $B$-fields smaller than $B_q = \frac{m^2 c^3}{e \hbar}
= 4.41\times 10^{13}$~G (Schwinger's critical $B$-field). }. The method
of characteristic surfaces or shock waves followed here (introduced by
Hadamard (1903) \cite{HAD}) can be applied to any field theory having
hyperbolic field equations, including electrodynamics.  Hence, the
result in Eq.(\ref{FG-metric}) (see appendix), implies two possible
paths of propagation or polarization modes, according to the double
solution $\Omega_{\pm}$. Using Eq.(\ref{H-E}) to compute the Lagrangean 
derivatives $ L_{F}, L_{FF}, L_{GG}$, one arrives to the couple of 
effective metrics (first order in $\mu $)


\begin{equation}
{g}^{\mu\nu}_{\rm eff+} = g^{\mu\nu} + 8\mu F^{\mu\alpha}F_{\alpha}\mbox{}^{\nu}\; ,
\hskip 0.3 truecm g^{\mu\nu}_{\rm eff-} = g^{\mu\nu}+14\mu F^{\mu \alpha} F_{\alpha}\mbox{}^{\nu}\; .
\end{equation}

In terms of electric ($E$) and magnetic ($B$) fields the tensor $F_{\mu\nu}$
can be written as

\begin{equation}
F_{\mu\nu} = E_{\mu}V_{\nu}-E_{\nu}V_{\mu} + \eta_{\mu\nu}\mbox{}^{\alpha\beta}
V_{\alpha}B_{\beta},
\end{equation}

where $V^{\mu}$ is the normalized $(V^{\mu}V_{\mu}=1$, Eq.(\ref{SCHW}))
velocity of the reference frame where the fields are measured. Our main
concern here is the behavior of photons, in the NLED context, emitted
from a highly magnetized neutron star. Corotating charges in the pulsar
magnetosphere or a rotating magnetic dipole lead to induced $E$-fields
in the star surface.  We consider here slowly rotating neutron stars in
order that the $E$-field contribution could be neglected. Since
$E^\alpha = 0$, the above expression simplifies to: $F_{\mu\nu} =
\eta_{\mu\nu} \mbox{}^{\alpha\beta} V_{\alpha} B_{\beta}$. Its
self-product reads

\begin{equation}
F^{\mu \alpha} F_{\alpha}\mbox{}^{\nu} = -
B^{\mu}B^{\nu} - B^2 \left(g^{\mu\nu} - V^{\mu}V^{\nu}\right) \;
\label{eq80} ,
\end{equation}

where $B^2 = B^\mu B_\mu$. After discarding a nonphysical conformal factor 
the effective metrics become

\begin{equation}
g_{\mu\nu}^{\rm eff_{\pm}} = g^{\mu\nu}-\frac{{\tilde{\beta}}^2_{\pm}}
{1 - {\tilde{\beta}}^2_{\pm} B^2} B^{\mu} B^{\nu} + \frac{{\tilde{\beta}}^2_{\pm} B^2} {1-{\tilde{\beta}}^2_{\pm}B^2} 
V^{\mu}V^{\nu} \; . 
\label{contrav-metric}
\end{equation}

The inverse (covariant) metric of Eq.(\ref{contrav-metric}) is obtained from 
the relation $ g_{\mu\nu} g^{\nu\alpha} = \delta_{\mu }\mbox{}^{\alpha}\;$, 
which then reads

\begin{equation}
g_{\mu\nu}^{\rm eff_{\pm}} = g_{\mu\nu} + {\tilde{\beta}}^2_{\pm} B^{\mu} 
B^{\nu} - {\tilde{\beta}}^2_{\pm} B^2 V^{\mu} V^{\nu}\; .
\end{equation}

By ${\tilde{\beta}}^2_{\pm}$ we mean $\tilde{\beta}^2_{+}=14\mu$ and 
$\tilde{\beta}^2_{-} = 8\mu$. In order to pursue our calculations we
assume a radial magnetic field. This is clearly not realistic, but is
an approximate description of the field geometry near the polar caps
of a magnetized neutron star. Thus

\begin{equation}
B_{\mu}=|B|l_{\mu}, \hskip 0.3 truecm l_{\mu}=\sqrt{-g_{rr}} \, \, 
\delta^{r}_{\mu},\hskip 0.3 truecm V_{\mu}=\sqrt{g_{00}}  \, \, 
\delta^{0}_{\mu}\; , \label{4-vectors}
\end{equation}

and consequently one arrives to the effective metric components

\begin{equation}
g^{\rm eff}_{00_\pm} = (1-{\tilde{\beta}}^2_{\pm}B^2)~g_{00} \;, \hskip 0.3 truecm 
g^{\rm eff_\pm}_{rr} = (1-{\tilde{\beta}}^2_{\pm}B^2)~g_{rr}\; . \label{l-dependent}
\end{equation}

Notice the structural similarity of both metric components.  From
Eq.(\ref{l-dependent}) is straightforward to rewrite the expression for
the apparent radius of the spherical star in Eq.(\ref{R-MAX}) in the
putative background metric given by Eq.(\ref{SCHW}). After doing so,
one can compute the ratio of the star area with H-E NLED effects
included to the area without (Einstein's derivation), as seen from the
distant observer. This quantity is represented here by $ {\cal{N}}
\label{area-ratio}$. \footnote{ We stress here that the NLED effects
depend on the effective geometry, which in turn depends upon the
strength of the magnetic field. Although the magnetic field $B$ is a
solution of the NLED field equations, to compute the ratio
"${\cal{N}}$" is only needed to know the local (on the star surface)
magnitude or modulus of the field. In the particular case of the
computation of the null geodesics discussed below, is indeed necessary 
to know the analytic expression for the field as a function of the 
coordinates, in particular of the radial $r$-coordinate. Unfortunately, 
such a solution for the magnetic field structure (or dependence on the coordinates) is only known for the linear (Maxwell's) case. }

\begin{figure}[htb]
{\epsfxsize=6.6cm\epsfbox{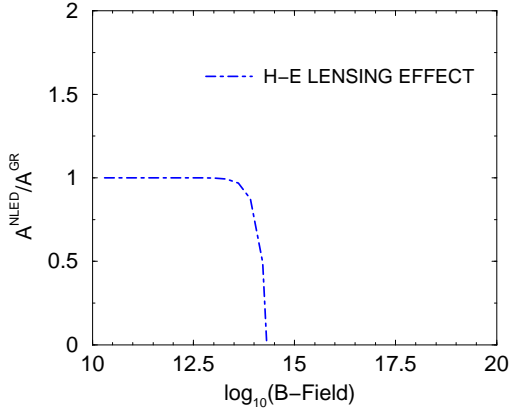} } 
\caption{H-E NLED to GR area ratio, as seen from a distant observer, for an
arbitrary $M/R$ ratio. Notice that the ratio grows down to zero near the QED 
critical field $B \sim 4.4 \times 10^{13}$~G. This means that no visible area 
is seen from the SMP. This ``anomalous'' behavior invalidates the Denisov 
et al.  results; which advogate for crucial changes in using the H-E approximation (see discussion below).}
\end{figure}


\subsubsection{ Qualitative description of the visible (apparent) star 
surface}

As stated above, "${\cal{N}}$" is intended to be a quantity describing
the ratio between the area of the star as seen by an observer at
infinity that receives the actual traveling luminous ray, and the
actual physical area of the neutron star, both projected on the plane
of the sky. This ratio of areas can be consistently figured out by the
following qualitative reasoning.

In a general relativistic framework, the local neutron star luminosity 
is defined as

\be
L_\gamma = 4 \pi R^2_\star \sigma_{\rm SB} T_\star^4 ,
\ee

where $R_\star$ is the neutron star radius and $T_\star$ is its temperature.
Both measured locally at the star surface. 
The luminosity measured by the observer at infinity, which carries the 
information on the spacetime metric around the neutron star, is defined 

\begin{equation}
L_\infty = L_\gamma \left(1 - \frac{R_s}{R_\star}\right) ,
\end{equation}

where {$\left(1 - \frac{R_s}{R_\star}\right)$} is the $tt$-metric component 
of a Schwarzschild spacetime , i.e., 

\begin{equation}
g_{tt} = \left(1 - \frac{R_s}{R_\star}\right) .
\end{equation} 

Notice that $g_{rr}  = \frac{1}{g_{tt}}$. On the other hand, the luminosity 
at infinity can be written as

\begin{equation} 
L_\infty = 4 \pi R^2_\infty \sigma_{\rm SB} (T_\star^\infty)^4 .
\end{equation}

Thus, the temperature measured by an observer at infinity relates to the 
actual star temperature via 

\begin{equation}
T_\star^\infty = T_\star \left(1 - \frac{R_s}{R_\star}\right)^{1/2} .
\end{equation}

Therefore, after manipulating all these relations one arrives to the relation 
between both the star radii (local and at infinity) 

\begin{equation}
R_\infty = {R_\star} {\left(g_{rr}\right)^{1/2} } .
\end{equation} 

In an equivalent approach, once one realizes that the star luminosity is
modified by the spacetime geometry induced by the matter distribution, one 
can readily verify that in the case of NLED its geometric effects on the 
photon propagation from the  ultra magnetized neutron star leads to an 
expression that is similar to the one above. Hence, one can write, taking 
into account the general relativistic effects superposed to the NLED effects, 
the local neutron star luminosity as given by

\begin{equation}
L_\gamma = 4 \pi R^2 \sigma_{\rm SB} T_\star^4 .
\end{equation}

The luminosity measured by the observer at infinity, which carries the 
information on the spacetime metric as modified by NLED, as in the 
Heisenberg-Euler (H-E) NLED, now reads

\begin{equation}
L^{eff}_\infty = L_\gamma \left(1 - \frac{R_s}{R_\star}\right)  
\left[1 - {\tilde{\beta}}^2_{\pm}B^2\right] 
\end{equation}

or equivalently,

\begin{equation}
L_\infty^{eff}  =  4 \pi ({R^{eff}_\infty})^2 \sigma_{\rm SB} 
(T_\star^\infty)^4 ,
\end{equation}

where the term $\left(1 - \frac{R_s}{R_\star}\right) \left[ 1 - {\tilde{\beta}}^2_{\pm} B^2 \right]$ now substitutes the $tt$-metric component of the pure Schwarzschild spacetime presented above. Notice that in this case the effective $tt$-metric component reads

\begin{equation}
g_{tt}^{eff} = \left(1 - \frac{R_s}{R_\star}\right) \left[1 - {\tilde{\beta}}^2_{\pm} B^2 \right]~ .
\end{equation} 

 This implies that 

\begin{equation}
g_{rr}^{eff}  = {( g_{rr} )}{ \left[ 1 - {\tilde{\beta}}^2_{\pm} B^2 \right] }  
=  \frac{ \left[  1 - {\tilde{\beta}}^2_{\pm} B^2 \right]}{ g_{tt} } .
\end{equation}  

Thus, the temperature measured by an observer at infinity now reads

\begin{equation}
T_\star^\infty = T_\star \left(1 - \frac{R_s}{R_\star}\right)^{1/2} 
\left[{1 - {\tilde{\beta}}^2_{\pm} B^2 } \right]^{1/2} ~ .
\end{equation}

Once again, after manipulating these relations one obtains

\begin{equation}
R_\infty^{eff} = {R_\star} \left({ g_{rr}^{eff} }\right)^{1/2} .
\end{equation} 

From these relations one can write the area ratio "${\cal{N}}$" that estimates 
the magnitude of the change in the apparent visible area of the star as seen 
from infinity compared to the actual star area, as measured by a local 
observer :   

\be
{\cal{N}} = \frac{\rm Area_{local}}{ \rm Area^{NLED}_{\infty}} = \frac{4 \pi (R_\star)^2 }{4 \pi (R^{eff}_\infty)^2 }.
\ee

Hence, by substituting the corresponding terms in these last equations one 
arrives to 

\be
{\cal{N}} = \left(\frac{g_{rr}}{ g_{rr}^{eff} }\right) = \frac{1}{\left[ 1 
- {\tilde{\beta}}^2_{\pm} B^2   \right]}~ . 
\ee 

As one can notice, in this new expression for the area ratio there is
no any functional of the $r$ coordinate but rather only a function of
the magnetic field strength at the surface of the neutron star. This 
is the reason of why we took $B$ as a constant in the numerical
calculation leading to the plots in Figures 1, and 2.



\subsection{Born-Infeld NLED}

The propagation of light from hypermagnetized neutron stars can also be 
viewed within the framework of the Born-Infeld Lagrangean

\begin{equation}
L= - b^2\left( \sqrt{1+\frac{F}{b^2}-G^2} -1\right)\label{exc}
\end{equation}

where $b^2 = \frac{e}{R_0^2} = {{e^4}/{m_0^2 c^8}} \longrightarrow b 
= 9.8 \times 10^{15}\; {\rm e.s.u.}$ As is well known, this is an
exceptional Lagrangean. One of its remarkable properties  is 
that it does not exhibit birefringence \cite{plebanski,Novelloal2000}.
In this case, the deduction we present in the appendix fails since the
quantities $\Omega_i$, with $ i=1,2,3$ ; vanish identically. We need
hence to go back to Eq.(\ref{eq25},\ref{eq26}); and substituting in
these equations the form for $L(F,G)$ of Eq.(\ref{exc}) to obtain a 
unique {\it characteristic equation}, which then leads to the effective 
metric

\begin{equation}
g^{\mu\nu}_{\rm eff} = g^{\mu\nu}+\frac{2}{b^2+F} F^{\mu}\mbox{}_{\alpha} 
F^{\alpha\nu}\; .
\end{equation}

Using the self-product of the tensor $F_{\mu\nu}$ given by (\ref{eq80}), 
and noting that in our case $F =  F_{\mu \nu} F^{\mu \nu} = 2 B^2$,
the effective metric then reads (see reference \cite{Novelloal2000} for
details)

\begin{equation} 
g^{\mu \nu}_{\rm eff}  =  g^{\mu \nu}  + \frac{2 B^2}{b^2} \left( V^\mu 
V^\nu -  l^\mu l^\nu \right)\; .
\end{equation}

By computing the inverse metric via $ g^{\mu \nu}_{\rm eff} g_{\nu
\alpha}^{\rm eff} = \delta^{\mu}_{{\;} {\;} \alpha}$, the covariant
form of this effective metric is obtained  as

\begin{equation}
g_{\mu \nu}^{\rm eff} =  g_{\mu \nu} - \frac{2B^2/b^2}{
(2B^2/b^2 + 1) } V_\mu V_\nu  + \frac{2B^2/b^2}{ (2B^2/b^2 + 1) }
l_\mu l_\nu \; . \label{effective-metric}
\end{equation}

After recalling the relations for $l_\nu $ and $B^\mu$ given in
Eq.(\ref{4-vectors}), one can verify that the covariant $rr$
effective metric component is then written as

\begin{equation}
g_{r r}^{\rm eff} = g_{r r} - \frac{2B^2/b^2}{ (2B^2/b^2 + 1) }~  g_{r r} =
  \left[\frac{1 }{ 1 + 2B^2/b^2 }\right]~ g_{r r} \; .
\end{equation}


\begin{figure*}[htb]
{\epsfxsize=6.6cm\epsfbox{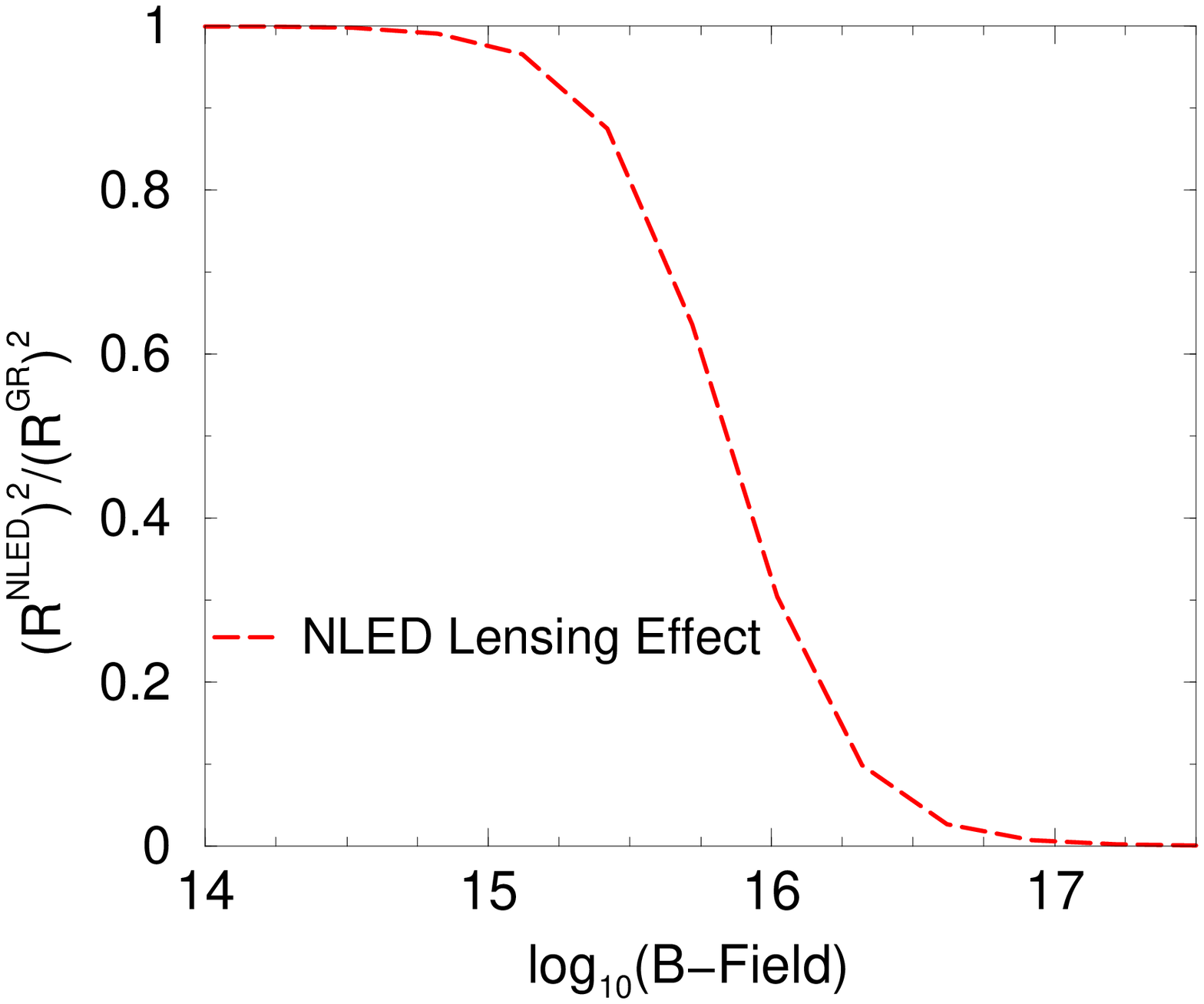} } 
{\epsfxsize=6.6cm\epsfbox{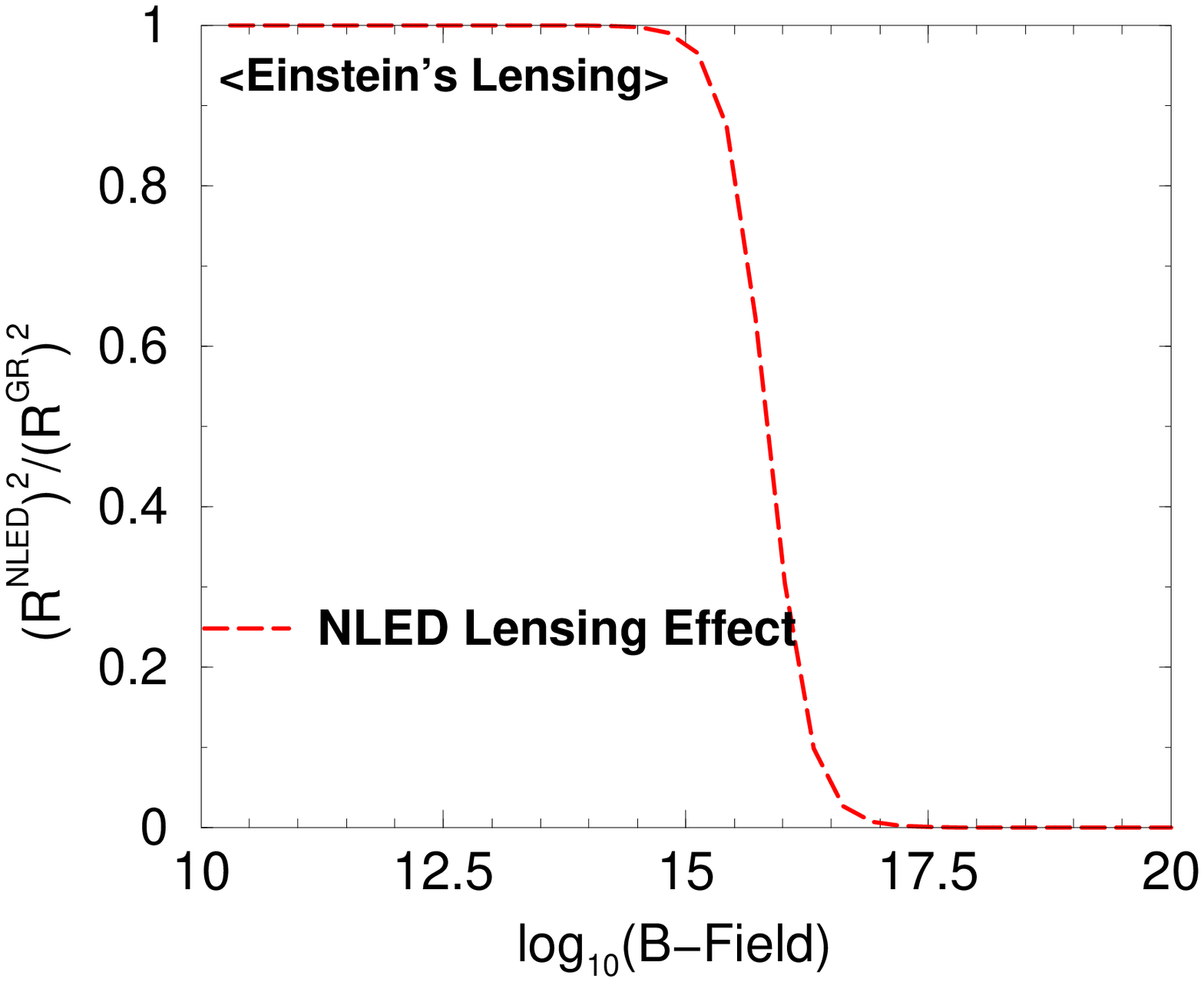} }  
\caption{H-E NLED to GR area ratio, as seen from a distant observer, for an
arbitrary $M/R$ ratio. Notice that the ratio grows down to zero near the QED 
critical field $B \sim 4.4 \times 10^{13}$~G. This means that no visible area 
is seen from the SMP.}
\end{figure*}

By following steps similar to the H-E case, the area ratio is presented
in Fig.2.  Although our hypothesis about the field geometry
maximizes the NLED effect on the observed properties of the star, they
should properly be taken into account when analyzing the X-ray
emission. (This point will be considered in a forthcoming paper). It
should also be relevant in pondering the effects on supernova dynamics
of photons radiated either by highly magnetized proto-neutron stars 
(Miralles et al. (2002) \cite{miralles2002}) or stellar-mass black 
holes enshrouded by super strong $B$-fields (see van Putten et al. (2004) \cite{vanputten2004}).


\section{Photon trajectories as seen from a distance} 

A self-consistent treatment of this problem does require the study of
the light propagation in the neutron star spacetime geometry modified
by NLED. In NLED the Lagrangean defined by Eq.(\ref{SCHW}) becomes 


\begin{eqnarray}
L(\dot{x}^\alpha, {x}^\alpha) & = & - \mu(B) \left(1 - \frac{R_{S}}{r}\right) 
\left(\frac{dt}{d\lambda}\right)^2 + \frac{\mu(B)}{ \left(1 - \frac{R_{S}}{r}\right) } \left(\frac{dr}{d\lambda}\right)^2 \nonumber \\
& + & r^2\left(\frac{d\theta}{d\lambda}\right)^2 + r^2\sin^2 
\theta \left(\frac{d\phi}{d\lambda}\right)^2 \; ,  
\end{eqnarray}


where $\mu(B) \equiv \frac{1 }{ 1 + 2B^2/b^2 } $. The resulting
Euler-Lagrange equations read (with $u^\alpha \equiv \frac{dx^\alpha}
{d\lambda} $; $\alpha = t,r,\theta,\phi$): 

\be
\frac{d}{d\lambda} \left(- 2 g^{\rm eff}_{tt} u^t \right) = 0  
\longrightarrow g^{\rm eff}_{tt} u^t = E_0 , \label{tt-component}
\ee 

\be
\frac{d}{d\lambda} \left(2 g^{\rm eff}_{\theta \theta} u^\theta \right) = 0  \longrightarrow g_{\theta \theta} u^\theta = L_0 , \label{thetatheta-component} 
\ee

\be
\frac{d}{d\lambda} \left(2 g^{\rm eff}_{\phi \phi} u^\phi \right) = 0
\longrightarrow g^{\rm eff}_{\phi \phi} u^\phi = \tilde{h} 
\label{phiphi-component} , 
\ee

where $\tilde{h}$ is defined as the {\it impact parameter} of the photon propagating from the NS surface to an observer at $r \longrightarrow \infty$, 
and $L_0$ a constant, which may be put equal to zero.
 
As the photons travel along null geodesics, the Euler-Lagrange equation for 
the $rr$-component of the metric can be obtained from the (first integral) 
relation

\begin{equation}
g^{\rm eff}_{tt} (u^t)^2 + g^{\rm eff}_{rr} (u^r)^2 + g_{\theta \theta} (u^\theta)^2 + g^{\rm eff}_{\phi \phi} (u^\phi)^2 \equiv 0  \; . 
\label{geodesics}
\end{equation}

In analogy with the planetary system one can chose a particular ``orbital'' plane at $\theta = \pi/2$, so that $u^\theta  = 0$ and $\sin^2 \theta = 1$, 
and thus $g^{\rm eff}_{\phi \phi} = r^2$. By making use of the standard 
change of variables ($u = 1/r$, $^\prime = d/dr$, and defining $A = 2 
B^2/b^2$) one can write the photon propagation equation as

\begin{equation}
E_0^2 - \left(\frac{1}{A + 1}\right)^2 \tilde{h}^2 (u^\prime)^2 - (1 
- R_S u) \left(\frac{1}{A + 1}\right) \tilde{h}^2 u^2 = 0 \; ,
\end{equation}

which one can in turn write as ($E_0$, $\tilde{h}$ constants) 

\begin{equation}
(u^\prime)^2 + u^2 = \frac{E_0}{\tilde{h}^2 } + R_S u^3 + \left(2 \frac{E_0} 
{ \tilde{h}^2 } - u^2 - R_S u^3\right)~A + \frac{E_0}{\tilde{h}^2 }~A^2 .
\label{photon-traject}
\end{equation}

After performing the derivative of equation (\ref{photon-traject}) with 
respect to the
angular coordinate $\phi$, and taking into account the fact that the
solution $u = constant$ can be discarted by an analogous (physical) 
reason as it is done in the case of a pure photon propagation in a 
Schwarzschild geometry, one arrives to

\begin{equation}
u^{\prime \prime} + (1 + A) u = \frac{3}{2} R_s \left(1 - A \right) u^2
+ \left[ 1 + R_s U - 2 \frac{E_0}{h} (1 + A) u^{-2} \right] \frac{dA}{dr} .
\label{nonlinear}
\end{equation}

This equation is highly nonlinear and quite hard to solve by analytical
means. (Notice that Eq.(\ref{nonlinear}) is quite similar to that one
obtains in the case of the photon propagation in a Schwarzschild
spacetime. Nonetheless, we stress that in order to find such a solution;
 an analogous procedure as the one usually pursued in general relativity 
({\it the perturbation approach}) does not work in the present case, since
for the extremely large magnetic field we are considering the induced
NLED effects are of the same order of magnitude as those purely
obtained from gravitation.) Besides, one needs to know the function 
$B(r, \phi)$, which is a solution of the Born-Infeld equations for this 
problem. Perhaps a numerical solution can be found,
but that would require extra work that we think would not add too much
to the new results we are presenting in this version of the paper. (A 
full detailed solution will be given elsewhere).

Eqs.(\ref{photon-traject},\ref{nonlinear}) shows very clearly that NLED
does modify the standard propagation of photons as compared to that one
in a pure Schwarzschild gravitational field \cite{vittorio01}, as described 
in most textbooks of GTR. A first look at Eq.(\ref{photon-traject}) suggests
that very large $B$-fields do reduce the effective star's visible area.



\section{Discussion and conclusion}

The nonlinear electrodynamics is a well stablished theory of the
electromagnetic interaction, which overcomes the failure of Maxwell's
theory in describing the stability of atomic structures. Since its
introduction in physics by Euler and Heisenberg \cite{H-E} it has
gained the status of a theory with a strong experimental support (for a
recent complete review see Delphenich (2003) \cite{Delphenich2003}). 
Despite being known since long, its very key features have become 
recognized only in the late years (see references above).
It had been shown in the last few years (Novello et al2000; 
Novello \& Salim 2001) that
the force acting along the photons trajectory can be geometrized  in
such a way that in an effective metric $ g^{eff}_{\mu\nu} = g_{\mu\nu}
+ g^{\rm NLED}_{\mu\nu}$ the photons follow geodesic paths in that
effective geometry (Novello et al2000; 
Novello \& Salim 2001) \cite{Novelloal2000,NS2001}.  In particular, the
effects of nonlinear electrodynamics in the physics of strongly
magnetized neutron stars have been recently studied by Mosquera Cuesta
\& Salim (2003; 2004) \cite{nos2004,nos2004a}. It was shown there that 
for extremely supercritical magnetic fields nonlinear electrodynamics
effects force photons to propagate along accelerated curves, in such a
way that the surface gravitational redshift of emission lines from
hypermagnetized neutron stars, as measured by a distant observer, is
significantly modified. Such a new effect turns out to be of the
same order of magnitude of the one produced by the pure gravitational
field.  

In this paper, we present a new astrophysical application of the
peculiar effects brought by NLED, which is intimately related to the
one previously discussed by (Mosquera Cuesta \& Salim 2003; 2004)
\cite{nos2004,nos2004a}. We show that the famous gravitational lensing
effect, originally introduced by Einstein, is also modified when the
electromagnetic (nonlinear) effects are taken into the depiction of the
physics around strongly magnetized neutron stars. Such a dynamics is
described by using both the Euler-Heisenberg and the Born-Infeld
approaches to NLED. Incidentally, we show, by comparing
Eqs.(\ref{l-dependent}) and (\ref{effective-metric}), that
Eq.(\ref{l-dependent}) clearly exhibits a divergence.  Such an
unphysical  behavior, in turn, invalidates the results presented by
Denisov et al. in Ref. \cite{Denisov} obtained within the H-E approach,
since they extended to the case of magnetars (neutron stars having
$B$-fields $\sim 10^{15}$~G) their formula to estimate the light-ray
bending angle beyond the QED $B$-field limit.  That is inconsistent.
Besides, both approaches: the H-E and the B-I NLED induce critical
changes in the $rr$ metric component that modifies the quantification
of the lensing effect. More relevant, yet, the area of a putative star
may appear to a distant observer nonphysically diminished in H-E NLED,
whereas it is physically largely reduced, and may even ``disappear'',
for fields $B \geq 10^{17}$~G in the B-I approach. The impact of this
on the star flux may be dramatic. Thence, the effect introduced by
(Mosquera Cuesta \& Salim 2003; 2004) \cite{nos2004,nos2004a} and this
new here crucially alters the dynamics of photons from very high
$B$-field stars, while entangles (makes it difficult) the inference of
their physical properties.

To the very end, what kind of observations can be performed in a search 
for this peculiar effect? As light is the prime messenger from the stars,
one can look for variability of very short period in the x-ray (or
$\gamma$-ray) light curves from most hypermagnetized pulsars. A
comparative analysis of the characteristics of those light curves with
the ones from canonical pulsars may render the clues on the presence of
NLED effects. A similar analysis can be pursued by using the
measurements of the surface gravitational redshift of that kind of
hypermagnetic stars.



\section{APPENDIX: The method of effective geometry}

Following Hadamard (1903) 
\cite{HAD}, the surface of discontinuity \footnote{ Of course, the
entire discussion onwards could alternatively be rephrased using
concepts more familiar to the astronomy community as that of light
rays used for describing propagation of EM waves in geometric
optics.} of the EM field is denoted by $\Sigma$. The field is
continuous when crossing $\Sigma$, while its first derivative
presents a finite discontinuity. These properties are specified as
follows: $\left[F_{\mu\nu}\right]_{\Sigma} =  0\; ,$ \hskip 0.3 truecm $\left[F_{\mu\nu|\lambda}\right]_{\Sigma} = f_{\mu\nu}k_\lambda \; \protect
\label{eq14} \;$ , where the symbol $\left[F_{\mu\nu}\right]_{\Sigma} = 
\lim_{\delta \to 0^+} \left(J|_{\Sigma + \delta}-J|_{\Sigma - \delta}\right)$
represents the discontinuity of the arbitrary function $J$ through the
surface $\Sigma$. The tensor $f_{\mu\nu}$ is called the discontinuity
of the field,  $k_{\lambda} = \partial_{\lambda} \Sigma $ is the
propagation vector, and the symbol ``$_|$'' stands for partial derivative.

Here-after we investigate the effects of nonlinearities of very strong
$B$-fields in the evolution of EM waves; described onwards as the
surface of discontinuity of the EM field (represented here-to-fore by
$F_{\mu\nu}$).  Extremizing the Lagrangian  with respect to the
potentials $A_{\mu}$ yields the following field equation: 

\be
\nabla_{\nu} (L_{F}F^{\mu\nu} + L_{G} F^{*\mu\nu}) = 0\label{eq60} . 
\ee

Besides this, we have the cyclic identity: 

\be
\nabla_{\nu}F^{*\mu\nu} = 0 \hskip 0.3 truecm \Leftrightarrow \hskip 0.3 truecm F_{\mu\nu|\alpha} + F_{\alpha\mu|\nu} + F_{\nu\alpha|\mu} = 0\; . \label{eq62}
\ee

The field equation can be written explicitly as: 

\be
L_{F}\nabla_{\nu}F^{\mu\nu} + 2 N^{\nu\mu\alpha\beta} \nabla_{\nu}F_{\alpha\beta}=0\label{eq61} \; ,
\ee

where the tensor $N$ is defined as

\begin{eqnarray}
N^{\mu\nu\alpha\beta}& = & L_{FF}F^{\mu\nu}F^{\alpha\beta} +
L_{FG}(F^{\mu\nu}F^{*\alpha\beta}
\\ \nonumber
&+& F^{*\mu\nu}F^{\alpha\beta}) +
L_{GG}F^{*\alpha\beta}F^{*\mu\nu} \label{eq20}\; .
\end{eqnarray}


Taking the discontinuities of the field equation we get:

\be
f_{\beta \lambda}k^\lambda = - \frac{2}{L_F} N_{\beta}^{\mu \nu \rho}
f_{\nu \rho} k_\mu .
\ee

The discontinuity of the  Bianchi identity yields: 

\be
f_{\alpha\beta}k_{\gamma} + f_{\gamma\alpha}k_{\beta} + f_{\beta
\gamma} k_{\alpha} = 0 . 
\ee

From these equations we obtain (see \cite{vittorio} for details)

\begin{eqnarray}
\chi k^2 & =
&\frac{4}{L_F}F^{\mu\nu}F^{\tau}\mbox{}_{\mu}k_{\nu}k_{\tau}(L_{FF}
\chi+L_{GF}\chi^{*}) \nonumber \\
& - & \frac{G}{L_F}k^2(L_{FG}\chi+L_{GG}\chi^{*}) \label{eq25}
\end{eqnarray}

\begin{eqnarray}
\chi^{*} k^2 & = &
\frac{4}{L_F}F^{\mu\nu}F^{\tau}\mbox{}_{\mu}k_{\nu}k_{\tau}
(L_{FG}\chi+L_{GG}\chi^{*}) - \frac{G}{L_F}k^2 \nonumber \\
& & (L_{FF}\chi + L_{FG}\chi^{*}) + \frac{2 F}{L_F}
k^2(L_{FG} \chi+L_{GG} \chi^{*}) ~ , \mbox{} \mbox{} \label{eq26}  
\end{eqnarray}

where we introduce the notation: $\chi=F^{\alpha\beta}f_{\alpha\beta}$,
$\chi^{*} = F^{*}_{\alpha\beta}f^{\alpha\beta}$, $k^2=g^{\mu\nu} k_{\mu}
k_{\nu}$. As the H-E QED Lagrangean is not a functional of the product 
$F*G$, then one obtains

\begin{eqnarray}
\chi k^2  =  \frac{4}{L_F}F^{\mu\nu}F^{\tau}\mbox{}_{\mu}k_{\nu}
k_{\tau} (L_{FF} \chi) -
\frac{G}{L_F}k^2(+L_{GG}\chi^{*})  \label{63}
\end{eqnarray}

\begin{eqnarray}
\chi^{*} k^2 & = & \frac{4}{L_F}F^{\mu\nu}F^{\tau}\mbox{}_{\mu}k_{\nu}
k_{\tau} (L_{GG} \chi^{*}) - \frac{G}{L_F}k^2(L_{FF}\chi) \nonumber \\
& + & \frac{2 F}{L_F}k^2(L_{GG} \chi^{*}) \label{64} \; .
\end{eqnarray}

We seek thus for a master relation representing the propagation of
field discontinuities, which should be independent of the quantities
$f_{\alpha \beta}$, that is, independent of $\chi$ and $\chi^{*}$.
There is a simple way to achieve such a goal. We firstly isolate the
common term $F^{\mu\nu} F^{\tau}\mbox{}_{\mu} k_{\nu}k_{\tau}$ which
appears in both Eqs.(\ref{63},\ref{64}). Then, by assuming that
$k^2\neq 0$, the difference of these equations can be put in the form
of an algebraic linear relation between $\chi$ and $\chi^{*}$ given as: 
$\Omega_{1} \chi^{*2} + \Omega_{2}\chi \chi^{*} + \Omega_{3}\chi^2 = 0,
\label{65}$ where we define: $\hskip 0.3 truecm \Omega_1 = \frac{G}{L_F} 
L^2_{GG}, \hskip 0.3 truecm $ and  $\Omega_2 = L_{GG} + \frac{F}{L_F} 
L_{FF} L_{GG}, \hskip 0.3 truecm$ and $\Omega_3 = - \frac{G}{L_F}L^2_{FF}$. 
Solving the quadratic equation for $\chi^*$ we obtain: $\chi^* = 
\Omega_{\pm} \chi$, with: $\Omega_{\pm} = \frac{-\Omega_{2\pm} \sqrt{
\Omega^2_2 - 4\Omega_1 \Omega_3}}{2\Omega_1} \label{polarizations}$. 
Using this solution in Eqs.(\ref{63}) and (\ref{64}), and assuming $\chi 
\neq 0$, after some algebra one gets the dispersion relation

\begin{equation}
\underbrace{ \left(g^{\mu\nu} - 4 \frac{L_{FF}} {L_F+G\Omega_{\pm} L_{GG}} 
F^{\lambda \mu}F^{\nu}\mbox{}_{\lambda}\right) }_{ g_{\mu \nu}^{\rm eff} } 
k_{\mu} k_{\nu} = 0\; . \label{FG-metric}
\end{equation}

Thence, one concludes that the discontinuities will follow geodesics in this
effective metric $g_{\mu \nu}^{\rm eff}$.

{Acknowledgements.-- JMS thanks CNPq (Brazil). HJMC thanks FAPERJ (Brazil).}

\end{document}